\documentclass[pra,twocolumn,amsmath,amssymb,showpacs, superscriptaddress,notitlepage,nofootinbib]{revtex4-2}
\usepackage{pdfpages} 
\usepackage{pgffor}

\usepackage[normalem]{ulem}

\usepackage{graphicx}
\usepackage[caption=false]{subfig}
\usepackage{booktabs}
\usepackage{mathtools,physics}
\usepackage[utf8]{inputenc}
\usepackage{enumitem}
\usepackage{appendix}

\usepackage[hidelinks]{hyperref} 
\hypersetup{
	colorlinks   = true, 
	urlcolor     = Blue, 
	linkcolor    = Blue, 
	citecolor   = Blue 
}
\usepackage[capitalize]{cleveref}

\usepackage [english]{babel}
\usepackage [autostyle, english = american]{csquotes} \MakeOuterQuote{"}

\usepackage[dvipsnames]{xcolor}

\newcommand{\blue}[1]{#1}

\newif\ifmargin

\marginfalse

\ifmargin
\newdimen\extramargin
\fi

\usepackage[author=Guillermo,color=yellow]{pdfcomment}

\makeatletter
\AtBeginDocument{\let\LS@rot\@undefined}
\makeatother

\def\supplementfilename{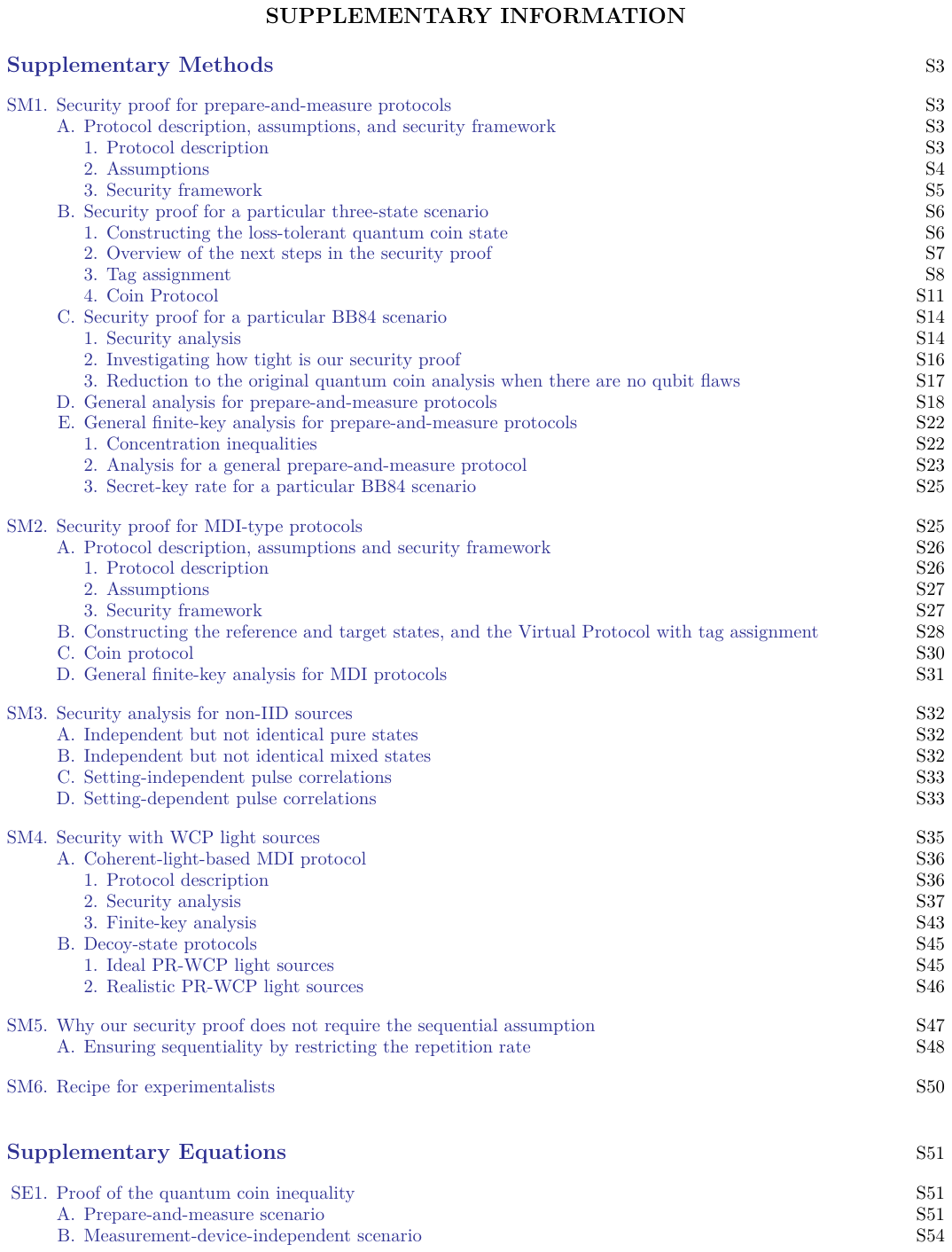}

\pdfximage{\supplementfilename}
\def\numbersupplementpages{\the\pdflastximagepages}

\newif\ifarXiv
\arXivtrue

\begin{document}

    \author{Guillermo Currás-Lorenzo} \email{gcurras@vqcc.uvigo.es}
    \affiliation{Faculty of Engineering, University of Toyama, Gofuku 3190, Toyama 930-8555, Japan}  
    \affiliation{Vigo Quantum Communication Center, University of Vigo, Vigo E-36310, Spain}  
    \affiliation{Escuela de Ingeniería de Telecomunicación, Department of Signal Theory and Communications, University of Vigo, Vigo E-36310, Spain}  
    \affiliation{atlanTTic Research Center, University of Vigo, Vigo E-36310, Spain}  

	\author{Margarida Pereira}	
    \affiliation{Faculty of Engineering, University of Toyama, Gofuku 3190, Toyama 930-8555, Japan}  
    \affiliation{Vigo Quantum Communication Center, University of Vigo, Vigo E-36310, Spain}  
    \affiliation{Escuela de Ingeniería de Telecomunicación, Department of Signal Theory and Communications, University of Vigo, Vigo E-36310, Spain}  
    \affiliation{atlanTTic Research Center, University of Vigo, Vigo E-36310, Spain}  

    \author{Go Kato}
    \affiliation{National Institute of Information and Communications Technology, Nukui-kita, Koganei, Tokyo 184-8795 Japan}

    \author{Marcos Curty}
    \affiliation{Vigo Quantum Communication Center, University of Vigo, Vigo E-36310, Spain}  
    \affiliation{Escuela de Ingeniería de Telecomunicación, Department of Signal Theory and Communications, University of Vigo, Vigo E-36310, Spain}  
    \affiliation{atlanTTic Research Center, University of Vigo, Vigo E-36310, Spain}  

	\author{Kiyoshi Tamaki} 
    \affiliation{Faculty of Engineering, University of Toyama, Gofuku 3190, Toyama 930-8555, Japan} 

     \title{Security framework for quantum key distribution with imperfect sources}

	\begin{abstract}
    {Imperfect {bit-and-basis} encoders compromise the security of quantum key distribution (QKD) systems via modulation flaws, side channels and inter-pulse correlations, which invalidate standard security proofs. Existing results addressing such imperfections suffer from critical limitations: they either consider only specific flaws, offer an unreasonably poor performance, or require the protocol to be run very slowly. Here, we present a finite-key security proof approach against coherent attacks that incorporates general {bit-and-basis} encoding imperfections (including modulation flaws, side channels and inter-pulse correlations) while achieving significantly better performances than previous approaches and requiring only partial characterization.} 
		
	\end{abstract}
	\maketitle

\section{Introduction} 

Quantum key distribution (QKD) can theoretically achieve the Holy Grail of cryptography, unconditional security against eavesdropping. However, in practice, discrepancies between the mathematical models assumed in security proofs and the actual functioning of the devices used in implementations prevent it from reaching this goal. Device-independent QKD \cite{mayersQuantumCryptography1998,barrettNoSignaling2005,acinDeviceIndependentSecurity2007} is currently not a satisfactory solution to this problem, as its performance is extremely poor \cite{nadlingerExperimentalQuantum2022,zhangDeviceindependentQuantum2022,liuPhotonicDemonstration2022,zapateroAdvancesDeviceindependent2023} and, in any case, its security proofs assume that the user devices leak absolutely no information to the outside \cite{zapateroImplementationSecurity2023,xuSecureQuantum2020}. On the other hand, measurement-device-independent (MDI) QKD \cite{loMeasurementDeviceIndependentQuantum2012} provides a {practical approach to} guarantee security with arbitrarily flawed receivers while achieving high performance. Thus, the remaining challenge is ensuring the security of QKD with {imperfect sources, such as sources that suffer from imperfections and side channels in the encoding of bit-and-basis information}.

So far, all efforts in this regard \cite{gottesmanSecurityQuantum2004,loSecurityQuantum2007,koashiSimpleSecurity2009,tamakiLosstolerantQuantum2014,pereiraQuantumKey2019,pereiraQuantumKey2020} have come at a price; some proofs are suitable only for particular encoding imperfections such as qubit flaws \cite{tamakiLosstolerantQuantum2014}, while others severely compromise the system's performance, i.e., its {repetition rate} \cite{pereiraQuantumKey2019,pereiraQuantumKey2020,metgerSecurityQuantum2023, sandfuchsSecurityDifferential2025} and {maximum achievable} distance \cite{gottesmanSecurityQuantum2004,loSecurityQuantum2007,koashiSimpleSecurity2009} {(see \textit{Discussion})}. Here, we overcome these crucial problems by presenting a security proof in the finite-key regime against coherent attacks that can incorporate {bit-and-basis} encoding imperfections and side channels while achieving much higher performances than previous approaches. \blue{Our approach unifies ideas from the quantum coin \cite{gottesmanSecurityQuantum2004,loSecurityQuantum2007,koashiSimpleSecurity2009} and loss-tolerant \cite{tamakiLosstolerantQuantum2014} security analyses in a way that naturally preserves their respective advantages---including their validity against the most general attacks allowed by quantum mechanics without any sequential restrictions---while overcoming their respective limitations.} Moreover, our proof requires only partial state characterization, which facilitates its application to real-life implementations.

\section{Results}

 \subsection{Partial state characterization assumption} 
 
 For ease of discussion, in the main text, we present our security proof approach by considering its application to an imperfect BB84 \cite{bennettQuantumCryptography1984} protocol; in the Supplementary Information, we provide a general description of our approach, and explicitly show how to apply it to other schemes. More precisely, {here} we consider that, {in each round $k$}, the sender (Alice) probabilistically selects a setting $j \in \{0_Z,1_Z,0_X,1_X\}$ and sends a signal in the state {$\rho_j^{(k)}$} to the receiver (Bob). The states {$\rho_j^{(k)}$} are partially characterized:\ they are known to be $\epsilon$-close (in terms of fidelity) to some characterized states $\{\ket{\phi_{j}}_{B}\}_j$, i.e.,{
	\begin{equation}
		\label{eq:emitted_state_assumption}
		 \ev{\rho_j^{(k)}}{\phi_j}_{B} \geq 1 - \epsilon,
	\end{equation}}for known $0\le\epsilon\le1$, which could more generally depend on $j$. In this discussion, for concreteness, we will assume that $\{\ket{\phi_{j}}_{B}\}_j$ is any set of qubit states, including flawed versions of the eigenstates $\ket{j}_{B}$ of the Pauli operators. {We remark, however, that $\{\ket{\phi_{j}}_{B}\}_j$ could more generally be \textit{any} set of characterized states (see Appendix \ref{subsec:more_general_assumption}).}

	We refer to the deviation between $\ket{\phi_{j}}_{B}$ and $\ket{j}_{B}$ as the qubit flaw, and to the deviation between {$\rho_j^{(k)}$} and $\ket{\phi_{j}}_{B}$ as the side channel. 
    \Cref{eq:emitted_state_assumption} is general in that it is not necessary to specify the cause for a non-zero $\epsilon$; it can cover {\textit{any}} kind of practical passive information leakage, such as those due to mode dependencies {\cite{nauerthInformationLeakage2009,xuExperimentalQuantum2015,gnanapandithanHiddenMultidimensional2025}}, electromagnetic or acoustic radiation \cite{gandolfiElectromagneticAnalysis2001}, or power consumption \cite{kocherDifferentialPower1999}; and \textit{any} kind of practical active information leakage, such as those due to Trojan-horse attacks (THAs) \cite{pereiraQuantumKey2019,pereiraQuantumKey2020,pereiraModifiedBB842023,navarretePracticalQuantum2021,mizutaniSecurityRoundrobin2021}. The partial state characterization assumed by our {security proof} is useful because some imperfections could {in principle live in arbitrarily high-dimensional spaces, making it extremely challenging to fully characterize them in practice.} 
    
    {In our main text discussion, for concreteness, we will consider that the emitted states are pure, and we will not mark their possible dependence on the round $k$, i.e., we will assume that $\rho_j^{(k)} = \ketbra{\psi_j}_B$. However, we remark that our security proof is directly applicable even if the emitted states are mixed and/or different for different rounds $k$, as long as \cref{eq:emitted_state_assumption} holds {(see Appendix \ref{subsec:mixed_non_identical_states})}.} We {also} remark that our security proof {is applicable even if the emitted states are correlated, although in this case, one needs to consider a modification to \cref{eq:emitted_state_assumption} and to the post-processing step of the protocol (see Appendix \ref{subsec:pulse_correlations})}.
	
	Due to \cref{eq:emitted_state_assumption}, the emitted states $\ket{\psi_{j}}_{B}$, which we denote by $\ket{\psi_{j}(\epsilon_j)}_{B}$ hereafter, can be expressed as {\cite{pereiraQuantumKey2019}}
	\begin{equation}
		\label{eq:emitted_states_assumed_form}
		{\ket{\psi_{j}(\epsilon_j)}_{B} =\sqrt{1-\epsilon_j} \ket{\phi_j}_{B} + \sqrt{\epsilon_j} \ket*{\phi_j^\perp}_{B}{,}}
	\end{equation}
	which is simply a state expansion in the basis $\{\ket{\phi_j}_{B}, \ket*{\phi_j^\perp}_{B}\}$. Here, $0\le\epsilon_j\le\epsilon$, and $\ket*{\phi_j^\perp}_{B}$ could be \textit{any} state orthogonal to $\ket{\phi_j}_{B}$, which implies that $\{\ket{\psi_{j}(\epsilon_j)}_{B}\}_j$ could be linearly independent when $\epsilon\neq0$. {Importantly, t}his fact makes the system inherently vulnerable to channel loss, as Eve could exploit it to enhance the distinguishability of the emitted states caused by the side channels. For example, {she} could perform an unambiguous state discrimination (USD) \cite{ivanovicHowDifferentiate1987} measurement and ensure that conclusive (inconclusive) events result (do not result) in detections. This implies that, if $\epsilon$ and the observed channel loss are high enough, Eve could have learned Alice's setting choices for all detected rounds without introducing any errors, and no security proof {can} provide {a} positive key rate. On the other hand, qubit flaws do not cause this vulnerability, as in their presence $\{\ket{\psi_{j}(0)}_{B}\}_j$ remain linearly dependent, and there is no USD measurement for linearly dependent qubit states.
	
	Side channels and qubit flaws are thus qualitatively different, and to achieve the best possible performance, security proofs should treat them differently. However, {so far,} no security proof achieves this satisfactorily.  The loss-tolerant (LT) analysis \cite{tamakiLosstolerantQuantum2014} deal{s} with qubit flaws tightly, and in fact show{s} that they have almost no impact on the secret-key rate, but cannot be applied in the presence of side channels, i.e., when $\epsilon > 0$. Conversely, the quantum coin analysis \cite{gottesmanSecurityQuantum2004,loSecurityQuantum2007,koashiSimpleSecurity2009} can be applied in the presence of both qubit flaws and side channels, but does not take into account their qualitative difference, and thus offers {an extremely} pessimistic performance for the former. Meanwhile, recent attempts to combine the strengths of both analyses, such as the {r}eference {t}echnique (RT) \cite{pereiraQuantumKey2020} (see also \cite{pereiraQuantumKey2019}), require an additional sequential assumption {\cite{pereiraModifiedBB842023,metgerSecurityQuantum2023,  sandfuchsSecurityDifferential2023}} that restricts the repetition rate at which the protocol can be run, severely reducing the secret-key rate obtainable in practice.
	
\subsection{Security analysis}
\label{sec:security_analysis}

 Here, we introduce a security proof approach that solves all of these drawbacks. To illustrate how, in this discussion, we consider a particular instance in which {the qubit components can be expressed as}
	{\begin{equation}
			\label{eq:qubit_state_model}
			\ket{\psi_{j}(0)}_B= \ket{\phi_{j}}_B =	\cos(\theta_j)\ket{0_Z}_{B}+\sin(\theta_j)\ket{1_Z}_{B},
		\end{equation}
		where {$\theta_j = (1+\delta/\pi) \varphi_j/2$,} $\varphi_j\in\{0,\pi,\pi/2,3\pi/2\}$ for $j\in\{0_Z,1_Z,0_X,1_X\}$, and $\delta \in [0,\pi)$ represents the magnitude of the qubit flaws. 
{We emphasize that this is just a specific example for illustration purposes, and that our general security proof, presented in the Supplementary Information, can be applied regardless of the specific form of the states $\{\ket{\phi_{j}}_B\}_j$.}
 {Moreover, without loss of generality (see Supplementary Information), we consider that $\epsilon_j=\epsilon$ for all $j$.} 
    The sifted key is generated from the detected key rounds, i.e., the rounds in which both Alice and Bob select the $Z$ basis and Bob obtains a bit value. In these rounds, instead of emitting $\ket{\psi_{0_Z}(\epsilon)}_{B}$ and $\ket{\psi_{1_Z}(\epsilon)}_{B}$ randomly, Alice could have generated the entangled state
	\begin{equation}
		\label{eq:Psi_Z_1}
		\ket{\Psi_Z (\epsilon)}_{AB}=  \frac{1}{\sqrt{2}} \big(\ket{0_Z}_A \ket{\psi_{0_Z} (\epsilon)}_B+ \ket{1_Z}_A \ket{\psi_{1_Z} (\epsilon)}_B\big)\,,
	\end{equation}
	and measured the ancillary system $A$ in the $Z$ basis. The amount of privacy amplification that needs to be applied to turn the sifted key into a secret key is directly related to the phase-error rate $e_{\rm ph}$, which is defined as the error rate that Alice and Bob would have observed if, in the detected key rounds, Alice had measured system $A$ in the $X$ basis {($X_A$)}, complementary to {the} $Z$ {basis}, and Bob had used {his actual} $X$-basis {measurement }{($X_B$)}. This is the case for security proofs based on the leftover hashing lemma with the entropic uncertainty relation \cite{rennerSecurityQuantum2005,tomamichelUncertaintyRelation2011,tomamichelTightFinitekey2012,curras-lorenzoTightFinitekey2021,tupkaryPhaseError2024}, and for those based on phase-error correction \cite{mayersQuantumKey1996,shorSimpleProof2000,koashiSimpleSecurity2009,hayashiConciseTight2012}, which have been shown to be essentially equivalent \cite{tsurumaruLeftoverHashing2020}.  
	
	The phase-error rate cannot be observed directly, and the goal of the security proof is to estimate it using the data obtained in the experiment. A common approach is to use the observed $X$-basis bit-error rate $e_{X}$. By noting that Alice could have replaced her $X$-basis emissions by the generation of 
	\begin{equation}
		\ket{\Psi_X (\epsilon)}_{AB}= \frac{1}{\sqrt{2}} \big(\ket{0_X}_A \ket{\psi_{0_X} (\epsilon)}_B - \ket{1_X}_A \ket{\psi_{1_X} (\epsilon)}_B \big),
	\end{equation}
	{one} can define $e_{\rm ph}$ and $e_{X}$ as the error rates {associated to} the {measurement of $X_A$ and $X_B$} on the detected rounds in which Alice prepares $\ket{\Psi_Z (\epsilon)}_{AB}$ and $\ket{\Psi_X (\epsilon)}_{AB}$, respectively. If $\delta$ and $\epsilon$ are close to zero, $\ket{\Psi_Z (\epsilon)}_{AB}$  and $\ket{\Psi_X (\epsilon)}_{AB}$ are close to each other, implying that $e_{\rm ph}$ and $e_{X}$ should also be relatively similar. This intuition was formalized in \cite{gottesmanSecurityQuantum2004,loSecurityQuantum2007,koashiSimpleSecurity2009} by introducing the quantum coin state 
	\begin{equation}
		\label{eq:quantum_coin_GLLP}
		\frac{1}{\sqrt{2}} \big(\ket{0_Z}_C \ket{\Psi_{Z} (\epsilon)}_{AB} + \ket{1_Z}_C \ket{\Psi_{X} (\epsilon)}_{AB} \big),
	\end{equation}
	and then considering that Alice probabilistically selects {one of the complementary bases} $Z_C$ or $X_C$ to measure the coin $C$. In particular, {these works showed} that the deviation between $e_{\rm ph}$ and $e_{X}$ can be bounded {by a function of} the fraction of events in which Alice obtained $X_C=1$ amongst the detected events in which she selected $X_C$. Of course, since {in the real protocol} Alice's coin is actually classical, this fraction cannot be observed. Instead, it can be upper bounded by considering the \textit{a priori} probability that Alice obtains $X_C =1$, given by $\frac{1}{2} \big(1-\textrm{Re}\braket{\Psi_{Z}(\epsilon)}{\Psi_{X}(\epsilon)}_{AB}\big)$, and assuming {the worst-case scenario in which} all events {such that $X_C =1$} are detected. However, the need to assume this scenario leads to a bound on $e_{\rm ph}$ whose tightness deteriorates as the channel loss increase{s}. The speed at which this occurs depends on the probability to obtain $X_C =1$, which grows with both $\delta$ and $\epsilon$. In the case of $\epsilon$, this behaviour is expected, as the effect of the side channels can be enhanced by Eve in the presence of loss. However, as previously discussed, this is not the case for qubit flaws, leading to a very loose bound when $\delta >0$. The solution to this {limitation} is the key to our security proof, and to understand it, it is helpful to first review the LT analysis, which {is} tight for $\delta >0$ but only valid when $\epsilon =0$.
	
	The main idea of the LT analysis is to consider the state resulting from the measurement of $X_A$ on $\ket{\Psi_Z (0)}_{AB}$ by re-expressing $\ket{\Psi_Z (0)}_{AB} = \sqrt{1-{q_0}} \ket{0_X}_A \ket{\psi_{\textrm{vir}0}(0)}_B + \sqrt{{q_0}} \ket{1_X}_A \ket{\psi_{\textrm{vir}1}(0)}_B$ with $q_0= \frac{1}{2} \big(1-\textrm{Re}\braket{\psi_{0_Z}(0)}{\psi_{1_Z}(0)}_{B}\big)$; we call $\ket{\psi_{\textrm{vir}\beta}(0)}_{B}$ ($\beta\in\{0,1\}$) the virtual states. That is, Alice emits $\ket{\psi_{\textrm{vir}0}(0)}_B$ ($\ket{\psi_{\textrm{vir}1}(0)}_B$) with probability $1-{q_0}$ (${q_0}$), and the estimation of the phase-error rate is reduced to the task of estimating the {$X_B$} detection statistics of these virtual states. {By using the fact that when $\epsilon=0$ the emitted states are {qubit states}, one can always find an operator form linear relationship between the actual states and the virtual states. {We remark that these linear relationships can always be derived regardless of the form of the qubit states $\{\ket{\psi_{j} (0)}_{B}\}_j$; the general procedure to do so can be found in \cite[Appendix~B]{curras-lorenzoFinitekeyAnalysis2021} and is discussed in our Supplementary Information.} When the qubit states have the particular form in \cref{eq:qubit_state_model}, one such  {linear} relationship is given by
		\begin{align}
			\label{eq:approx_0}
			&\ketbra{\psi_{\textrm{vir}0} (0)}_{B} = \ketbra{\psi_{0_X} (0)}_{B}, 
		\end{align}
		\begin{align}
			\label{eq:approx_1}
			&c_1 \ketbra{\psi_{0_Z} (0)}_{B} + \ketbra{\psi_{\textrm{vir}1} (0)}_{B} \nonumber \\
			&= c_2 \ketbra{\psi_{1_Z} (0)}_{B} +  c_3\ketbra{\psi_{1_X} (0)}_{B},
		\end{align}
		{where 
        	\begin{equation}
		\label{eq:c1_c2_c3}
	\begin{aligned}
		&c_1 \coloneqq \frac{\cos(\kappa \pi/2)}{\cos(\kappa \pi) - \cos(\kappa \pi/2)}, \\
		&c_2 \coloneqq \frac{\cos(\kappa \pi/2)}{\cos(\kappa \pi/2)-1}, \\
            &c_3 \coloneqq \frac{1 + \cos(\kappa \pi/2)}{\cos(\kappa \pi/2)-\cos(\kappa \pi)}.
	\end{aligned}
	\end{equation}
        with $\kappa = 1 + \delta/\pi$.} {The existence of such linear relationships} implies that the {$X_B$} detection statistics of the virtual states can be determined \textit{exactly} using the observed {$X_B$} detection statistics of the actual states{, including basis-mismatched events}. {As a result, the LT analysis allows a tight estimation of} $e_{\rm ph}$ even in the presence of high loss, and offers a key rate that is almost independent of $\delta$ when $\epsilon=0$.

	{The key idea of our proof to extend this behaviour to the case $\epsilon>0$ is to combine} the quantum coin analysis and the LT analysis through the notion of the target and reference states introduced {by} the RT \cite{pereiraQuantumKey2020}. Note that, when $\epsilon > 0$, the relationships in \cref{eq:approx_0,eq:approx_1} do not hold exactly, but {do} still hold approximately if $\epsilon \approx 0$, irrespectively of the value of $\delta$. {We} use this fact to construct a quantum coin state for which the probability to obtain $X_C=1$ is almost independent of $\delta$. 
	
	First of all, rather than considering $\ket{\Psi_Z (\epsilon)}_{AB}$, which is a purification of a convex combination of $\ketbra{\psi_{\textrm{vir}0} (\epsilon)}_{B}$ and $\ketbra{\psi_{\textrm{vir}1} (\epsilon)}_{B}$ taken according to the probabilities $1-{q_\epsilon}$ and $q_{\epsilon} = \frac{1}{2} \big(1-\textrm{Re}\braket{\psi_{0_Z}(\epsilon)}{\psi_{1_Z}(\epsilon)}_{B}\big)$; we consider instead a purification of a convex combination of the LHS of \cref{eq:approx_0,eq:approx_1} taken according to these probabilities (followed by normalization), i.e.,
	\begin{align}
		\label{eq:psi_tar}
		&\ket{\Psi_{\rm Tar} (\epsilon)}_{DAB} =\frac{1}{\sqrt{1+c_1 q_{\epsilon}}}\Big[ \sqrt{1-q_{\epsilon}} \ket{0}_D \ket{0}_A \ket{\psi_{\textrm{vir}0} (\epsilon)}_B \nonumber \\
		&+ \sqrt{q_{\epsilon}} \ket{1}_D \big(\sqrt{c_1}\ket{0}_A \ket{\psi_{0_Z} (\epsilon)}_B + \ket{1}_A \ket{\psi_{\textrm{vir}1} (\epsilon)}_B \big)\Big]{.}
	\end{align}
	Then, we define the target statistics as the error rate associated to the measurement $\{\ket{0}_D,\ket{1}_D\}$ and $X_B$ on the detected rounds after Eve's attack, i.e., the statistics of the outcomes ($\ket{0}_D$, $X_B=1$) and ($\ket{1}_D$, $X_B=0$). Note that, since this measurement commutes with the measurement $\{\ket{0}_A, \ket{1}_A\}$, the target statistics can be regarded as a mixture of the phase-error statistics and the statistics of the events in which Alice emits $\ket{\psi_{0_Z} (\epsilon)}_{B}$ and Bob obtains $X_B=0$, with the latter being observ{ed} in the actual protocol. Therefore, if we can estimate the overall target statistics, we can estimate the phase-error {rate}.
	
	To do so, we consider a purification of a similar convex combination of the RHS of \cref{eq:approx_0,eq:approx_1}, i.e.,
	\begin{align}
		\label{eq:psi_ref}
		&\ket{\Psi_{\rm Ref} (\epsilon)}_{DAB} = \frac{1}{\sqrt{1+c_1 q_{\epsilon}}}\Big[ \sqrt{1-q_{\epsilon}} \ket{0}_D \ket{0}_A \ket{\psi_{0_X} (\epsilon)}_B \nonumber \\
		&+ \sqrt{q_{\epsilon}} \ket{1}_D \big(\sqrt{c_2}\ket{0'}_A \ket{\psi_{1_Z} (\epsilon)}_B + \sqrt{c_3} \ket{1'}_A \ket{\psi_{1_X} (\epsilon)}_B \big)\Big],
	\end{align}
	where the orthonormal basis $\{\ket{0'}_A, \ket{1'}_A\}$ has been chosen such that $\ket{\Psi_{\rm Tar} (0)}_{DAB} =  \ket{\Psi_{\rm Ref} (0)}_{DAB}$, which is always possible due to the equality in \cref{eq:approx_1}. As before, we define the reference statistics as the error {rate} of the measurement $\{\ket{0}_D,\ket{1}_D\}$ and $X_B$ on the detected rounds after Eve's attack. Since this measurement commutes with the measurement $\{\ket{0'}_A, \ket{1'}_A\}$, the reference statistics can be regarded as a mixture of the statistics of the events in which Alice emits $\ket{\psi_{0_X} (\epsilon)}_{B}$ and Bob obtains $X_B=1$, and the events in which Alice emits $\ket{\psi_{1_Z} (\epsilon)}_{B}$ or $\ket{\psi_{1_X} (\epsilon)}_{B}$ and Bob obtains $X_B=0$, all of which are observ{ed}. In other words, the reference statistics can be determined using the data acquired in the actual protocol. 
	
	Finally, we define {the} loss-tolerant quantum coin state as
	\begin{align}
		\label{eq:quantum_coin}
		&\ket{\Psi_{\rm LT coin} (\epsilon)}_{CDAB} \nonumber \\
		&= \frac{1}{\sqrt{2}} \big(\ket{0_Z}_C \ket{\Psi_{\rm Tar} (\epsilon)}_{DAB} + \ket{1_Z}_C \ket{\Psi_{\rm Ref } (\epsilon)}_{DAB} \big),
	\end{align}
	and bound the deviation between the target and reference statistics by considering the probability to obtain $X_C=1$, given by   $\frac{1}{2} \big(1-\textrm{Re}\braket{\Psi_{\rm Tar}(\epsilon)}{\Psi_{\rm Ref}(\epsilon)}_{DAB}\big)$. When $\delta >0$, the resulting bound on the phase-error rate is much tighter than in the original quantum coin analysis, since $\textrm{Re}\braket{\Psi_{\rm Tar}(\epsilon)}{\Psi_{\rm Ref}(\epsilon)}_{DAB}$ is almost independent of $\delta$, while  $\textrm{Re}\braket{\Psi_{Z}(\epsilon)}{\Psi_{X}(\epsilon)}_{DAB}$ decreases rapidly as $\delta$ increases.

	To turn the above {argument} into a full security proof {against general attacks,} there remains a loose end to tie {up}. Unlike in the original quantum coin analysis, it is not possible to assume here that Alice replaces her actual source by the generation of  \cref{eq:quantum_coin} in all rounds, since the statistics of Alice's source differ in general from those of  \cref{eq:quantum_coin}. Instead, we {consider} that Alice randomly samples her emissions, where the sampling probabilities depend on her emitted state and are chosen such that a sampled {emission} is equivalent to {that originating from} \cref{eq:quantum_coin}. As shown in the Supplementary Information, this allows us to apply the above analysis to estimate the {number of phase errors} within the sampled rounds, and then extend this estimate to all rounds via known statistical results{, thus obtaining a bound on the overall phase-error rate that is valid even in the finite-key regime}. We note that {the} proof merely relies on the idea that Alice \textit{could} {in principle} sample her emissions; in the actual experiment, {however,} Alice does not actually need to {perform this sampling step, {n}or to} decide which sampled rounds correspond to measuring the quantum coin in the $Z_C$ or $X_C$ bases. The latter is another improvement over the original quantum coin analysis, which require{s} Alice to actually assign each round of the protocol to either the $Z_C$ or $X_C$ basis, and discard the data of the rounds assigned to $X_C$ \cite{wangFinitekeySecurity2018}.

    	{A critical advantage of our security proof is that it allows the protocol to be run at high repetition rates, unlike some previous approaches that address source imperfections. In particular, the RT \cite{pereiraQuantumKey2020}, while sharing many of the advantages of our approach (including the ability to incorporate side channels and resilience to qubit flaws), requires the assumption that the probability that Alice selects a particular bit and basis choice in round $k$ must be independent of Bob's previous $k-1$ measurement outcomes. This independence condition can only be guaranteed if Eve's attack is sequential, that is, if she is prevented from correlating Bob's measurement outcomes with Alice's setting choices in later rounds. In practice, enforcing this sequential condition requires running the protocol slowly such that Alice's emitted pulses could not possibly have influenced Bob's previous outcomes, even if Eve wanted to correlate them.

In contrast, our security proof directly guarantees security against the most general attacks allowed by quantum mechanics without any sequential restrictions, thus enabling high-speed operation. The key technical insight is that our proof is based on bounding the deviation between the reference and target statistics by considering that Alice generates the quantum coin state in \cref{eq:quantum_coin}, and then deriving a quantum coin inequality through the application of the Bloch sphere bound to each coin system $C$. To derive this inequality, we consider that Eve performs a coherent attack on all rounds simultaneously, after which \blue{Bob performs basis-independent quantum non-demolition measurements to determine which rounds are detected, and then Alice and Bob measure the systems corresponding to the detected rounds}. Crucially, the coin system $C$ corresponding to any particular \blue{detected} round $k$ always remains a qubit system, even when conditioning on Eve's global attack\blue{, on Bob's detection outcomes for all rounds,} and on all of Alice's and Bob's previous measurement outcomes in \blue{detected} rounds $1$ through $k-1$. As a result, \blue{we obtain a quantum coin inequality that relates the expectation values for the outcomes on} round $k$ \blue{conditional on} any outcomes in rounds 1 through $k-1$. \blue{We then apply concentration inequalities such as Azuma's inequality or Kato's inequality \cite{katoConcentrationInequality2020} to relate the conditional expectations to the actual statistics observed in the protocol. Importantly, all our proof steps hold} regardless of any correlations Eve might introduce between rounds.

We remark that, while in this section we have focused on the prepare-and-measure BB84 protocol for clarity, our security proof approach is broadly applicable to other QKD protocols as well. For example, it can be directly extended to standard MDI-QKD \cite{loMeasurementDeviceIndependentQuantum2012}. Since MDI-QKD eliminates all detector side channels while our proof addresses transmitter imperfections, combining these approaches enables security against both source and measurement device imperfections. The extension to MDI-QKD follows naturally because the bit-and-basis encoding imperfections at each transmitter can be incorporated using the same loss-tolerant quantum coin technique described above. Furthermore, our approach can be applied to other protocols, including three-state protocols and even an MDI-type protocol in which the users send non-phase-randomized coherent states \cite{navarretePracticalQuantum2021}, demonstrating the versatility of our techniques. For the full analysis of these scenarios, see Supplementary Information.

    \section{Discussion} 
    
    Now, we apply our loss-tolerant quantum coin analysis to evaluate the secret-key rate obtainable for BB84-type protocols in the presence of both qubit flaws and side channels, and discuss our findings by drawing a comparison with previous analyses. For the simulations, we assume the following parameters: error correction inefficiency $f = 1.16$, detector dark count probability $p_d = 10^{-8}$ \cite{xuSecureQuantum2020,pittaluga600kmRepeaterlike2021}, detector efficiency $\eta_d = 0.73$ \cite{pittaluga600kmRepeaterlike2021}, and a repetition rate of 2.5 GHz \cite{boaronSecureQuantum2018}.
	
	In \cref{fig:ourresults}, we plot the achievable secret-key rate in bits per second for both the BB84 and three-state protocols when using our analysis. We consider  $\delta = 0.063$, following the experimental results reported in \cite{honjoDifferentialphaseshiftQuantum2004,xuExperimentalQuantum2015}, and several values of $\epsilon$.  For both protocols, the key rate is sensitive to the value of $\epsilon$, which is expected, as higher values of $\epsilon$ make it easier for Eve to discriminate the emitted states in the presence of channel loss. When $\epsilon=0$, both protocols offer the same secret-key rate, as already known \cite{tamakiLosstolerantQuantum2014}. However, {consistently with previous results \cite{pereiraModifiedBB842023}, we find that,} for $\epsilon >0$, the additional state emitted in the BB84 protocol results in {a tighter phase-error rate estimation}, which translates to higher key rates.
		\begin{figure}[h]
			\includegraphics[width=8.5cm]{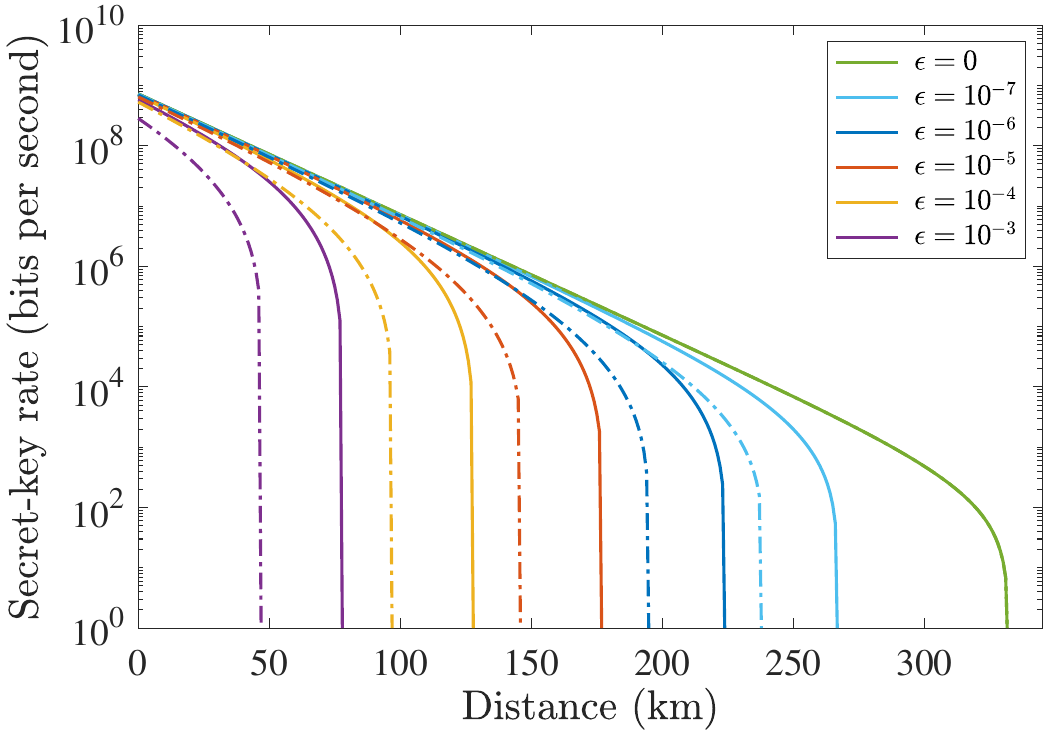} 
			\caption{{{Asymptotic} secret-key rate obtainable using our proof as a function of the distance (km) for the BB84 (solid lines) and three-state (dashed-dotted lines) protocols. We assume $\delta=0.063$ \cite{honjoDifferentialphaseshiftQuantum2004,xuExperimentalQuantum2015} and consider several values of $\epsilon$.}}
			\label{fig:ourresults}
		\end{figure}
		
		In \cref{fig:comparison}, we compare the secret-key rate obtainable using our proof with that of the original quantum coin analysis \cite{gottesmanSecurityQuantum2004,loSecurityQuantum2007,koashiSimpleSecurity2009} and the RT analysis \cite{pereiraQuantumKey2020,pereiraModifiedBB842023}. We consider the BB84 protocol, since the original quantum coin analysis cannot provide any key for the three-state protocol. Also, we fix $\epsilon = 10^{-6}$, and consider the presence ($\delta = 0.063$) and absence ($\delta = 0$) of qubit flaws. When there are no qubit flaws, our proof converges to the original quantum coin analysis. However, in their presence, the secret-key rate offered by our proof decreases {only} very slightly, while that of the original quantum coin analysis decreases dramatically. As already mentioned, applying the RT requires running the protocol sequentially, which limits the repetition rate. Therefore, for the RT, rather than using 2.5 GHz, we determine the maximum repetition rate {under the restriction} that Alice only emit{s} a pulse after Bob has finished his measurement of the previous pulse. For this calculation, we assume a standard fibre in which photons travel at {about} 2/3 of the speed of light \cite{agrawalFiberOpticCommunication2021}. As can {be} see{n} in \cref{fig:comparison}, this significantly restricts the secret-key rate achievable using the RT after the first few kilometers. We note that this sequential condition, which is also imposed by security proofs based on the generalized entropy accumulation theorem \cite{metgerSecurityQuantum2023,  sandfuchsSecurityDifferential2023}, is not required by our security proof{, as discussed in \cref{sec:security_analysis}}. In the Supplementary Information, we explain at length why this is the case.

	  \begin{figure}
		\includegraphics[width=8.5cm]{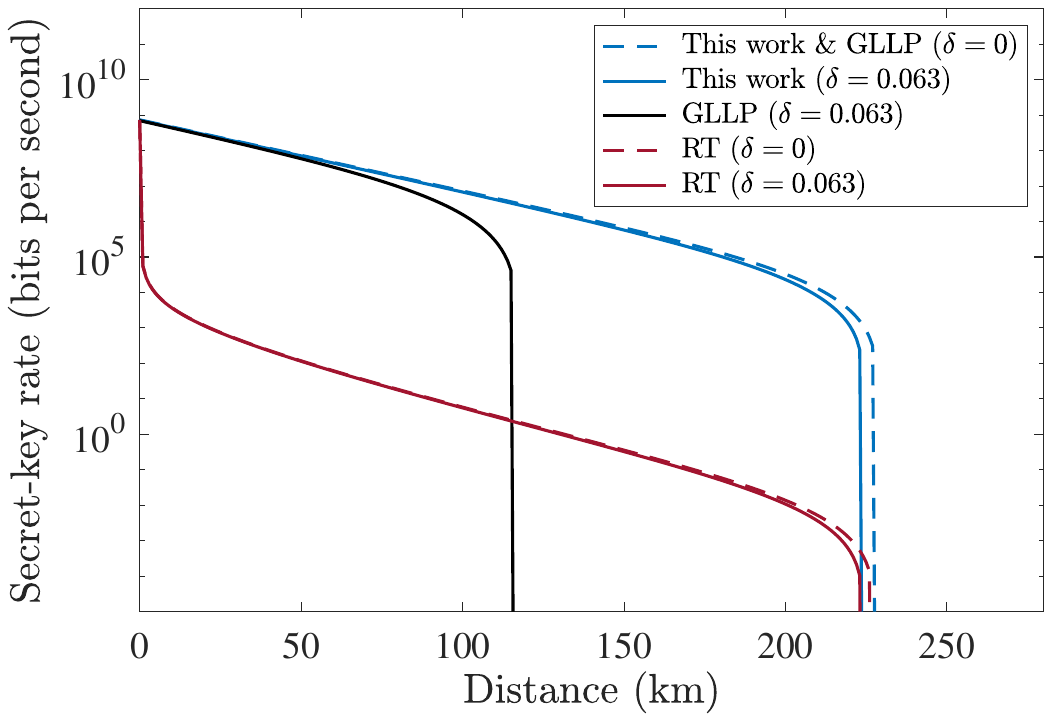} 
		\caption{{{Asymptotic} secret-key rate obtainable using our proof as a function of the distance (km) for the BB84 protocol, compared with that of the original quantum coin (GLLP) \cite{gottesmanSecurityQuantum2004,loSecurityQuantum2007,koashiSimpleSecurity2009} and {r}eference {t}echnique (RT) \cite{pereiraQuantumKey2020,pereiraModifiedBB842023} analyses. We consider the values $\epsilon = 10^{-6}$ and  $\delta \in \{0,0.063\}$.}}
		\label{fig:comparison}
	\end{figure}

\vspace{1cm}
    {In \cref{fig:ourresults,fig:comparison}, we have assumed the asymptotic regime in which the total number of emitted pulses, $N$, approaches infinity. However, our security {proof} can be applied to secure practical QKD implementations with a finite $N$. In \cref{fig:finite}, we show the achievable secret-key rate for several values of this parameter. {For} {large} $N$, the performance {approaches} that {of} the asymptotic regime.}
    \begin{figure}[h]
		\includegraphics[width=8.5cm]{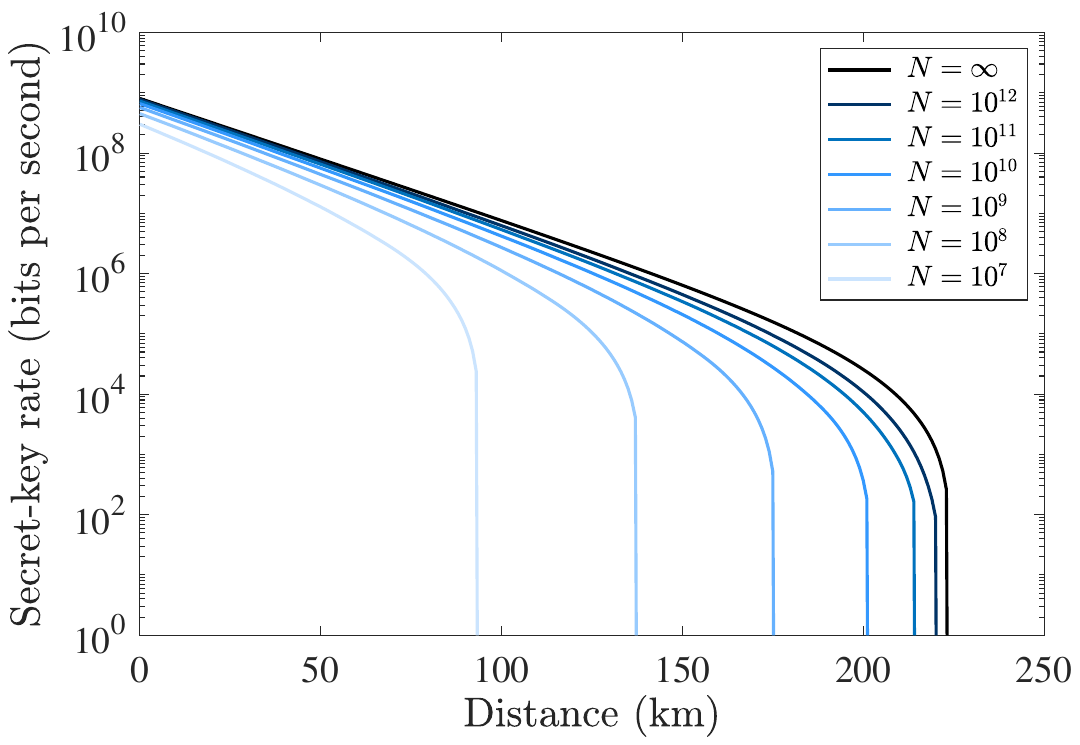} 
		\caption{{{Finite-size} secret-key rate against general attacks obtainable {using our proof} as a function of the total number of emitted pulses $N$. We consider the BB84 protocol, $\epsilon = 10^{-6}$, $\delta=0.063$, and we set the correctness and secrecy parameters of the final key to $\epsilon_{\rm corr} = \epsilon_{\rm secr} = 10^{-10}$.}}
		\label{fig:finite}
	\end{figure}

    \section{Conclusion}

We have introduced a security proof approach that can ensure the finite-key security of QKD protocols against the most general attacks allowed by quantum mechanics (i.e., coherent attacks) in the presence of bit-and-basis encoding imperfections. As \cref{fig:comparison} demonstrates, our analysis achieves significantly higher secret-key rates than previous results in the presence of both side channels and qubit flaws, guaranteeing the security of practical QKD setups without compromising their performance. Moreover, it does not need any characterization of the side channels, other than an upper bound on their overall magnitude. Importantly, when applied to the BB84 protocol, the asymptotic performance of our security proof is essentially optimal given its partial characterization assumptions \cite{curras-lorenzoNumericalSecurity2025}.

While our security proof represents a significant advance in addressing implementation imperfections while relaxing the need for full characterization, several important challenges remain for achieving truly secure QKD in practice. On the experimental front, accurately determining the parameter $\epsilon$ that bounds the magnitude of side channels remains a difficult task. Real QKD systems may suffer from multiple side channels simultaneously---including mode dependencies, setting-dependent correlations, electromagnetic radiation, acoustic emissions, power consumption variations, and susceptibility to Trojan-horse attacks---each requiring careful characterization, and new unexpected side channels are still being discovered \cite{gnanapandithanHiddenMultidimensional2025}. The total $\epsilon$ must account for all these leakage channels combined, and obtaining tight bounds requires sophisticated measurement techniques with very high accuracy and extremely low noise. Furthermore, as we have shown, if $\epsilon$ is too large, the system becomes vulnerable to USD attacks that can completely compromise security. This places stringent requirements on experimental implementations to minimize side-channel leakage through careful engineering, shielding, and isolation techniques. Modulator-free \cite{paraisoModulatorfreeQuantum2019} and passive QKD \cite{wangFullyPassive2023,zapateroFullyPassive2023} represent a promising approach to eliminate the side channels introduced by active components, and our techniques have been adapted to incorporate residual side channels in such setups \cite{navarreteSecurityPractical2024}.

On the theoretical front, while our proof addresses bit-and-basis encoding imperfections, extending it to simultaneously handle imperfections in decoy-state implementations remains an open challenge. Decoy-state QKD is crucial for practical implementations using weak coherent pulses, but decoy-state modulators can introduce their own side channels (including correlations in intensity modulation and phase randomization) that are not fully covered by our current analysis. Although we discuss in the Supplementary Information how our techniques could potentially be extended to address some of these issues, and simple imperfections like intensity fluctuations can be readily incorporated, a comprehensive treatment of general decoy-state imperfections—particularly in the finite-key regime—requires further theoretical development.

Despite these challenges, \blue{the techniques we have developed demonstrate considerable flexibility and broad applicability. While we have focused on the prepare-and-measure BB84 protocol for clarity in the main text, our security proof approach can be extended to many other QKD protocols. For example, it can be applied to MDI-QKD \cite{loMeasurementDeviceIndependentQuantum2012}. Since MDI-QKD eliminates all detector side channels while our proof addresses transmitter imperfections, combining these approaches enables security against both source and measurement device imperfections. The extension to MDI-QKD follows naturally because the bit-and-basis encoding imperfections at each transmitter can be incorporated using the same loss-tolerant quantum coin technique. Furthermore, our approach can be applied to other protocols, including three-state protocols and even an MDI-type protocol in which the users send non-phase-randomized coherent states \cite{navarretePracticalQuantum2021}, demonstrating the versatility of our techniques (see Appendix \ref{app:methods} and Supplementary Information for detailed analyses of these scenarios).}  Moreover, as shown in \cite{curras-lorenzoSecurityQuantum2025}, our security proof can be combined with the result in \cite{tupkaryPhaseError2024} to incorporate also detection efficiency mismatches in Bob's setup, going beyond previous attempts at simultaneously addressing source and detector imperfections \cite{sunSecurityQuantum2021,marcominiLosstolerantQuantum2024} in terms of generality and applicability in the finite-key regime. Beyond QKD, our techniques may also prove useful for other information-theoretic tasks in which information leakage must be addressed.

In short, we believe this work represents an important step toward bridging the gap between the theoretical promise of unconditional security and the practical reality of QKD implementations.

\appendix

\section{METHODS}
\label{app:methods}

{\subsection{Mixed and non-identically-distributed states}
\label{subsec:mixed_non_identical_states}

Our analysis directly applies even if the emitted states $\rho_j$ are mixed. To see why, note that, if \cref{eq:emitted_state_assumption} holds, by Uhlmann's theorem, there must exist a purification $\ket{\psi_j}_{BS}$ of $\rho_j$ such that
\begin{equation}
\label{eq:purification_assumption}
    \abs{\braket{\phi_j}{\psi_j}_{BS}}^2 = \ev{\rho_j}{\phi_j}_{B}  \geq 1-\epsilon,
\end{equation}
where we have defined $\ket{\phi_j}_{BS} \coloneqq \ket{\phi_j}_B \ket{0}_S$. This implies that our analysis for pure states can be directly applied to the mixed state case simply by substituting $\ket{\psi_j}_{B} \to \ket{\psi_j}_{BS}$ and $\ket{\phi_j}_{B} \to \ket{\phi_j}_{BS}$ throughout.

Also, our security proof can be applied without any modification even if the emitted states $\rho_j^{(k)}$ are different for different rounds $k$, as long as \cref{eq:emitted_state_assumption} holds for all rounds.  Note that, in this case, the target and reference states $\ket*{\Psi_{\rm Tar }^{(k)}(\epsilon)}$ and $\ket*{\Psi_{\rm Ref}^{(k)}(\epsilon)}$, and thus also the quantum coin state $\ket*{\Psi_{\rm LTcoin}^{(k)}(\epsilon)}$, depend on the round $k$. However, this is not a problem since one can use exactly the same procedure as in the case of identically distributed states to find a lower bound on $\textrm{Re} \braket*{\Psi_{\rm Ref}^{(k)}(\epsilon)}{\Psi_{\rm Tar}^{(k)}(\epsilon)}$ that holds for all rounds $k$, and apply our security proof as is.}

  \subsection{Determining the value of $\epsilon$}
  \label{subsec:determining_epsilon}
  
  {Any device used in a communication system can leak partial information about its internal settings to the channel. This means that any cryptosystem, including device-independent QKD \cite{mayersQuantumCryptography1998,barrettNoSignaling2005,acinDeviceIndependentSecurity2007}, necessarily requires a degree of characterization of such potential side channels to guarantee security. For this characterization, our security proof requires only a bound on the overall combined magnitude of all side channels --- the parameter $\epsilon$. That is, unlike other security {proofs} {(e.g., \cite{lucamariniPracticalSecurity2015})}, we do not need any state characterization of the side channels, which might be impossible to obtain in practice. Remarkably, this means that our proof could serve as a guideline to secure the source while significantly relaxing the need for detailed characterization. While obtaining a rigorous bound on $\epsilon$ for a given implementation is a non-trivial experimental problem that is outside the scope of this work, here, we discuss how one could combine information about various side channels in order to obtain a value for $\epsilon$.}
 
	One critical side channel is that caused by a THA, in which Eve injects light into Alice's source and then measures the back-reflected light to learn information about Alice's setting choice. The amount of {leaked} information can be related to the intensity of the back-reflected light, $\mu_{\rm out} = \gamma \mu_{\rm in}$, where $\mu_{\rm in}$ is the intensity of the injected light, and $\gamma$ represents the optical isolation of the transmitting unit. In particular, it is straightforward to show that the back-reflected light can be expressed as
	\begin{equation}
		\label{eq:tha}
		\ket{\xi_j}_E = \sqrt{1-\epsilon_j} \ket{v}_E + \sqrt{\epsilon_j} \ket{\Omega_{j}}_E,
	\end{equation}
	where  $\epsilon_j \leq \mu_{\rm out}$ \cite{pereiraModifiedBB842023}. Here, $\ket{v}_E$ is a vacuum state independent of {Alice's setting $j$}, and $\ket{\Omega_{j}}_E$ is a non-vacuum state that can in general depend on $j$. It has been argued \cite{lucamariniPracticalSecurity2015} that, for any given implementation, one can determine a threshold $\mu_{\rm in}^{\rm U}$ above which the injected light is very likely to damage the optical components of Alice's source and be detected. Based on this, one can obtain a bound $\mu_{\rm out}^{\rm U} \coloneqq \gamma \mu_{\rm in}^{\rm U}$ that can be reduced by adjusting the optical isolation $\gamma$. If Eve's THA is the only side channel present, the emitted states are $\{\ket{\phi_j}_B \otimes \ket{\xi_j}_E\}_j${, where $\{\ket{\phi_j}_B\}_j$ are qubit states. Therefore, we can apply our proof by considering} the set of qubit states $\{\ket{\phi_j}_B \otimes \ket{v}_E\}_j$ and setting $\epsilon = \epsilon_{\rm THA} \coloneqq \mu_{\rm out}^{\rm U}$. Note that, unlike other analyses \cite{lucamariniPracticalSecurity2015}, our proof does not need any assumption {on} Eve's injected light, such as it being a coherent state, other than a bound on its intensity. 
	 
	Beyond THAs, Alice's source may also passively leak information through unwanted modes. For example, consider polarization mode dependencies in phase-encoding setups. While the encoded pulses should ideally maintain constant polarization (e.g., horizontal), imperfect alignment between laser and phase modulator can cause the polarization to depend slightly on the setting choice $j$. The generated pulse then becomes
	\begin{equation}
		\label{eq:polarization_mode_dep}
		\ket{\psi_j}_B = \sqrt{1-\epsilon'_j} \ket{\phi_j}_{B_h} \ket{v}_{B_v} +  \sqrt{\epsilon'_j}  \ket{v}_{B_h} \ket{\phi_j}_{B_v},
	\end{equation} 
	where $B_h$ ($B_v$) denotes the horizontally (vertically) polarized mode. While the exact values of $\{\epsilon'_j\}_j$ may fluctuate, obtaining an upper bound $\epsilon'_j \leq \epsilon_{\rm MD}$ should be experimentally feasible. If necessary, its value may be reduced using countermeasures such as polarizing beam splitters. In the presence of both the THA and the polarization mode dependencies, the emitted states are $\{\ket{\psi_j}_B  \ket{\xi_{j}}_E\}_j$, and one can apply our analysis by defining the qubit states $\{\ket{\phi_j}_{B_h} \ket{v}_{B_v} \ket{v}_E\}_j$, which satisfy 
    \begin{equation}
    \!\abs{~_{B_h} \!\!\bra{\phi_j} \!\! ~_{B_v}\!\! \bra{v} \!\! ~_E \! \bra{v} \, \ket{\psi_j}_B  \ket{\xi_{j}}_E}^2 \leq \epsilon,
    \end{equation}
	where $\epsilon \coloneqq 1- (1-\epsilon_{\rm THA})(1-\epsilon_{\rm MD}) \leq \epsilon_{\rm THA} + \epsilon_{\rm MD}$. That is, the magnitudes of the side channels simply combine additively. Other forms of information leakage, such as electromagnetic/acoustic radiation, or temporal/spectral/spatial mode dependencies, can also be written in the form of \cref{eq:tha,eq:polarization_mode_dep}, and thus their individual magnitudes also contribute additively towards the overall $\epsilon$.  {Pulse correlations can also be essentially regarded as a form of information leakage and incorporated into $\epsilon$ in this way (see below).}

Based on all the above, we consider the following to be a promising approach: (1) identify the principal source side channels affecting a particular implementation; (2) obtain an upper bound on their magnitude;  (3) if necessary, apply countermeasures to reduce this magnitude; and (4) sum all the individual upper bounds. Side channels that are too small to be precisely quantified could be accounted for by conservatively increasing $\epsilon$.

\subsection{Pulse correlations}
\label{subsec:pulse_correlations}

Pulse correlations are a special type of side channel that occurs when the state emitted in the $k$-th round, denoted as $\ket*{\psi_{j_{k}|j_{k-1},j_{k-2},...,j_{k-l_c}}}_{B_{k}}$, depends not only on the $k$-th setting choice $j_k$, but also on the previous $l_c$ setting choices $j_{k-1},j_{k-2},...,j_{k-l_c}$, where $l_c$ denotes the maximum correlation length. In this case, these states should satisfy
\begin{equation}
\big\vert\braket*{\phi_{j_k}}{\psi_{j_k|j_{k-1}, j_{k-2}, ..., j_{k-l_c}}}_{B_k}\big\vert^2 \geq 1-\epsilon_{\rm qubit},
\end{equation}
and
\begin{align}
&\big|\braket*{\psi_{j_k|j_{k-1},...,\tilde{j}_{k-l},...,j_{k-l_c}}}{\psi_{j_k|j_{k-1},...,{j}_{k-l},...,j_{k-l_c}}}_{B_k}\big|^2 \nonumber\\
&\geq 1 - \epsilon_{l},
\end{align}
where $\{\ket*{\phi_{j}}_{B_k}\}_j$ is a set of known qubit states, $l \in \{1,...,l_c\}$, and $\tilde{j}_{k-l}$ is a setting choice that differs from $j_{k-l}$. Then, one can apply our security proof by setting
\begin{equation}
\epsilon = 1- (1-\epsilon_{\rm qubit})(1-\epsilon_{\rm correl}) \leq \epsilon_{\rm qubit} + \epsilon_{\rm correl}
\end{equation}
where 
\begin{equation}
\epsilon_{\rm correl} \coloneqq 1- \prod_{l=1}^{l_c} (1-\epsilon_l) \leq \sum_{l=1}^{l_c} \epsilon_l.
\end{equation}
 Also, one needs to divide the protocol rounds into $(l_c+1)$ groups according to the value of $k \bmod (l_c+1)$, and apply post-processing separately for each group {\cite{pereiraQuantumKey2020,mizutaniSecurityRoundrobin2021,pereiraModifiedBB842023}}. While this does not affect the asymptotic key rate, it can have an impact in the finite-key regime, as the blocksize is effectively reduced from $N$ to $N/(l_c+1)$ rounds. Also, although this discussion implicitly assumes a finite maximum correlation length $l_c$, our security proof can also be applied when the correlations have an unbounded length, as recently shown in \cite{pereiraQuantumKey2024a} (see also \cite{agulleiroModelingCharacterization2025}). For more details, see the Supplementary Information.

\subsection{A more general form of Eq.~(1)}
\label{subsec:more_general_assumption}

{In the main text, we have assumed that the emitted states $\rho_j^{(k)}$ are close in fidelity to some known qubit states $\{\ket*{\phi_j}_B\}_j$, see \cref{eq:emitted_state_assumption}.} However, our security proof can be applied under a more general assumption; namely,
    \begin{equation}	\label{eq:emitted_state_assumption_nonqub}
		 \ev*{\rho_j^{(k)}}{\tilde \phi_j}_B \geq 1 - \epsilon,
    \end{equation}
where $\{\ket*{\tilde \phi_j}_B\}_j$ are \textit{any} characterized states, not necessarily qubits. This is because, in our analysis, the fictitious qubit states $\{\ket*{\phi_j}_B\}_j$ only serve as a blueprint to appropriately define the states $\ket{\Psi_{\rm Tar}}_{DAB}$ and $\ket{\Psi_{\rm Ref}}_{DAB}$. The secret-key rate obtainable primarily depends on $\textrm{Re}\braket{\Psi_{\rm Tar}}{\Psi_{\rm Ref}}_{DAB}$, which is a linear combination of the inner products ${\rm Re}\braket{\psi_{j'}}{\psi_{j}}_B$ $\forall j,j'$.  Although these inner products are not known precisely, in the Supplementary Information, we show that, for any $\{\ket*{\tilde \phi_j}_B\}_j$, the problem of finding the minimum value of $\textrm{Re}\braket{\Psi_{\rm Tar}}{\Psi_{\rm Ref}}_{DAB}$ that is consistent with \cref{eq:emitted_state_assumption_nonqub} reduces to a numerically solvable semidefinite program. Our numerical simulations focus on the case $\ket*{\tilde \phi_j}_B = \ket*{\phi_j}_B$, for which \cref{eq:emitted_state_assumption_nonqub} becomes \cref{eq:emitted_state_assumption}. As explained in "Determining the value of $\epsilon$" above, this corresponds to minimal side-channel characterization, in which one knows only the overall magnitude of the side channels, but has no information on their specific form. While this minimal requirement is a key strength of our proof, given the experimental challenges of obtaining precise side-channel characterization, our proof is not \textit{only} applicable in this scenario. Namely, if one is able to obtain a partial characterization of some side channels, one could include this information into the definition of the states $\{\ket*{\tilde \phi_j}_B\}_j$. This would generally improve the bound on $\textrm{Re}\braket{\Psi_{\rm Tar}}{\Psi_{\rm Ref}}_{DAB}$, potentially resulting in a significant performance improvement for many practical side channels.

\subsection{Characterizing qubit flaws}
    
    In addition to $\epsilon$, {the} proof also requires knowledge of the qubit states $\{\ket{\phi_j}\}_j$,
	which can be acquired by testing the source. Methods to characterize this imperfection have been proposed and implemented, see e.g., \cite{xuExperimentalQuantum2015,huangCharacterizationStatePreparation2023}. In practice, these characterization tests could be subject to small inaccuracies. The simplest way to take these into account would be to incorporate any deviation between the estimated $\{\ket{\phi_j}\}_j$ and the actual qubit components of the emitted states into the parameter $\epsilon$. However, this is in general pessimistic, since Eve cannot enhance the effect of this deviation in the presence of loss. 
    
    A tighter approach would be to define $\{\ket{\phi_j}\}_j$ as the actual qubit components of the emitted states. This requires a slight modification to our security proof, as the states $\{\ket{\phi_j}\}_j$ would no longer be fully characterized. Consequently, the coefficients in \cref{eq:approx_0,eq:approx_1}, which appear in the definition of the target and reference states in \cref{eq:psi_ref,eq:psi_tar}, are no longer known precisely. However, as shown in \cite{pereiraModifiedBB842023}, one can calculate the maximum possible range for these coefficients and find the worst case scenario within those ranges. For more information on this approach, we refer the reader to \cite{pereiraModifiedBB842023}.

\subsection{Protection against detector imperfections side channels}
 
 Our security {proof} is designed to protect QKD implementations from source imperfections. When applied to BB84---or other prepare-and-measure (P\&M) scenarios---our security proof's only requirement for Bob's measurement is that it satisfies the basis-independent detection efficiency condition, i.e., that the probability that he obtains a successful bit outcome is independent of his choice of basis. While this requirement relaxes the need for an exact characterization of Bob's setup and can tolerate certain imperfections (such as the measurement bases not being mutually unbiased), {it is still a stringent condition that is only met in practice if all of Bob's detectors have the same efficiency. However, we remark that, as recently shown in \cite{curras-lorenzoSecurityQuantum2025}, our security proof can be readily combined with the result in \cite{tupkaryPhaseError2024} to incorporate detection efficiency mismatches.}
 
 {Still,} {detector control attacks \cite{lydersenHackingCommercial2010,gerhardtFullfieldImplementation2011,makarovControllingPassively2009,wiechersAftergateAttack2011}} constitute a significant threat to the security of BB84 and other P\&M protocols, and no solution at the security proof level is known to comprehensively deal with these \cite{xuSecureQuantum2020,zapateroImplementationSecurity2023}.  That being said, our {approach} is compatible with MDI-QKD \cite{loMeasurementDeviceIndependentQuantum2012}, which eliminates all detector-related security vulnerabilities by delegating measurements to an untrusted intermediary node. When applied to MDI protocols, our {analysis} secures both Alice's and Bob's sources against general imperfections, providing robust protection against both source and detector side channels. For details on applying our {security proof} to MDI-type protocols, see Supplementary Information.

	
\subsection{Weak coherent sources}

    Our security {proof} can also be applied when the users emit intensity-modulated phase-randomized weak coherent pulses, rather than single photons. In this case{, assuming ideal intensity modulation and phase randomization,} the emitted signals can be regarded as a statistical mixture of photon-number states, and our proof is directly applicable to obtain a bound on the number of phase errors within the single-photon events. The quantities needed to evaluate this bound are not directly observable, since the users do not know which events correspond to single photons. Nevertheless, one can simply obtain bounds for these quantities using the decoy-state method \cite{hwangQuantumKey2003,loDecoyState2005,wangBeatingPhotonNumberSplitting2005}.

    {For decoy-state QKD protocols, our security proof can directly incorporate general imperfections and side-channels in the bit-and-basis encoder, which addresses a major practical security concern. {Our proof can also straightforwardly incorporate fluctuations in the intensity modulation, since in their presence the emitted states can still be regarded as a mixture of photon-number states.} For scenarios with imperfect intensity modulation {(beyond simple fluctuations)} and/or imperfect phase randomization, additional considerations are needed since the emitted signals may no longer be perfectly described as a statistical mixture of photon-number states independent of the intensity choice. As explained in the Supplementary Information, the techniques introduced in our work could find applications in incorporating such decoy imperfections into security proofs while considering the finite-key regime and general attacks. This represents a promising research avenue, since previous works addressing such imperfections \cite{tamakiDecoystateQuantum2016,zapateroSecurityQuantum2021,sixtoSecurityDecoystate2022,curras-lorenzoSecurityQuantum2023,naharImperfectPhase2023,sixtoSecretKey2023,sixtoQuantumKey2025} have considered the asymptotic regime and/or collective attacks.}

    {An alternative approach to circumvent the security vulnerabilities associated with these imperfections is the {MDI-type} protocol proposed in \cite{navarretePracticalQuantum2021}, which employs coherent light but requires neither intensity modulation nor phase randomization. Our security {analysis} is directly compatible with this protocol, and offers significant enhancements in both security and performance compared with its original security proof \cite{navarretePracticalQuantum2021} based on the RT \cite{pereiraQuantumKey2020}. In particular, while the original proof is only valid against sequential attacks, ours can defend against the most general attacks allowed by quantum mechanics. {The combination of our {approach} with this protocol constitutes a practical solution that offers an unprecedented level of implementation security and performance.} For more details, including a comprehensive security analysis and numerical results, see the Supplementary Information.}

     \subsection{Choice of target and reference states} 
 
		We note that \cref{eq:psi_ref,eq:psi_tar} are not the only possible choice for $\ket{\Psi_{\rm Tar} (\epsilon)}_{DAB}$ and $\ket{\Psi_{\rm Ref} (\epsilon)}_{DAB}$. Essentially, these states need to satisfy two conditions: (1) when $\epsilon = 0$, $\ket{\Psi_{\rm Tar} (0)}_{DAB} = \ket{\Psi_{\rm Ref} (0)}_{DAB}$, as this ensures the resilience of the phase-error rate bound against qubit flaws; (2) the coefficient of the state $\ket{\psi_{\textrm{vir}0} (\epsilon)}_B$ ($\ket{\psi_{\textrm{vir}1} (\epsilon)}_B$) inside $\ket{\Psi_{\rm Tar} (\epsilon)}_{DAB}$ should be proportional to $\sqrt{1-q_\epsilon}$ ($\sqrt{q_\epsilon}$), as in $\ket{\Psi_Z (\epsilon)}_{AB}${, which is needed to ensure that phase errors are correctly defined.} To compute the results in \cref{fig:ourresults,fig:comparison,fig:finite}, we have used a slightly different choice of target and reference states than that in \cref{eq:psi_ref,eq:psi_tar}. The reason why this different choice is advantageous is explained in the Supplementary Information.

     {\subsection{Comparison with side-channel-secure QKD}

     Both our work and side-channel-secure (SCS) QKD \cite{wangPracticalLongDistance2019} aim to address source imperfections in QKD systems, but they take fundamentally different approaches. The main idea behind SCS-QKD is to design a variant of twin-field QKD \cite{lucamariniOvercomingRate2018} that is inherently immune to mode dependencies by having each user send only two states: a vacuum state and a non-vacuum state. Since the vacuum state is "mode independent", the protocol becomes immune to mode dependencies in the sense that, regardless of which mode the non-vacuum state occupies, this does not allow Eve to better distinguish between the two states. However, despite its name, SCS-QKD is not immune to other side channels that leak key information through systems different from the intended signal, such as unintended electromagnetic radiation, back-reflected light due to Trojan-horse attacks, or correlations between consecutive pulses. 

    Recently, a refined version of the protocol was proposed \cite{jiangSidechannelSecurity2024} to address a limitation of the original proposal---the requirement to generate perfect vacuum states---by demanding only a lower bound on the fidelity between the two emitted states. Although this is not discussed in \cite{jiangSidechannelSecurity2024}, we believe this new assumption could be exploited to incorporate side channels beyond mode dependencies (including pulse correlations through the approach described in Appendix \ref{subsec:pulse_correlations}) after bounding their overall magnitude, by introducing a parameter $\epsilon$ as in our work. Thus, certain aspects of our work, such as our state characterization (see Appendix \ref{subsec:determining_epsilon}), may be relevant for SCS-QKD as well.

     A limitation of SCS-QKD is that its key idea is fundamentally restricted to protocol designs where each user sends only two different states. In contrast, our work develops general techniques to incorporate source imperfections into security proofs of QKD, and is thus broader in scope. Our state characterization can incorporate general imperfections directly into the state used in the security proof, making it highly useful for the security analysis of QKD protocol in general. Although we focus mainly on BB84-type protocols, in the Supplementary Information, we demonstrate the versatility of our techniques by applying them to a MDI-type scenario in which the users send non-phase-randomized coherent states \cite{navarretePracticalQuantum2021}. This latter scenario differs from SCS-QKD mainly in that three states are emitted rather than two, and thus the approach we take to prove security is necessarily different.} 
    

\vspace{0.5cm}

\section*{Acknowledgements}
	{We thank Koji Azuma, Akihiro Mizutani, {Álvaro Navarrete and Víctor Zapatero} for valuable discussions.} {This work was supported by the {Galician Regional Government (consolidation of research units: atlanTTic), the Spanish Ministry of Economy and Competitiveness (MINECO), the Fondo Europeo de Desarrollo Regional (FEDER) through the grant No. PID2020-118178RB-C21, MICIN with funding from the European Union NextGenerationEU (PRTRC17.I1) and the Galician Regional Government with own funding through the “Planes Complementarios de I+D+I con las Comunidades Autonomas” in Quantum Communication, the “Hub Nacional de Excelencia en Comunicaciones Cuanticas” funded by the Spanish Ministry for Digital Transformation and the Public Service and the European Union NextGenerationEU, the European Union’s Horizon Europe Framework Programme under the Marie Sklodowska-Curie Grant No. 101072637 (Project QSI), the project “Quantum Security Networks Partnership” (QSNP, grant agreement No 101114043) and the European Union via the European Health and Digital Executive Agency (HADEA) under the Project QuTechSpace (grant 101135225).}. {G.C.-L.\ and M.P.\ acknowledge support from JSPS Postdoctoral Fellowships for Research in Japan.} G.C.-L. acknowledges funding from the European Union's Horizon Europe research and innovation programme under the Marie Skłodowska-Curie Postdoctoral Fellowship grant agreement No.\ 101149523. G.K. acknowledges support from JSPS Kakenhi (C) No.20K03779 and 21K03388. {K.T. acknowledges support from JSPS KAKENHI Grant Numbers JP18H05237 and 23H01096, and JST-CREST JPMJCR 1671.}

 \section*{Author contributions}
 G.C.-L. identified the need for the research project, and K.T.\ conceived the fundamental idea behind the security proof, with help from all the authors. K.T., G.C.-L.\ and M.P.\ developed the majority of the security proof, with
 contributions from G.K.\ and M.C.  M.P. and G.C.-L.\ performed the calculations and the numerical simulations. G.C.-L., M.P.\ and K.T.\ wrote the manuscript, and all {the} authors contributed towards improving it and checking the validity of the results. 

\section*{Competing interests}
 The authors declare no competing interests.

 \section*{Data availability statement}
All data generated and analyzed during this study is available from the corresponding author on reasonable request.

	\bibliography{refs,refs_extra}

	 \clearpage

	 \onecolumngrid
	 \ifarXiv
	 \foreach \x in {1,...,\numbersupplementpages}
	 {
	 	\includepdf[pages={\x}]{\supplementfilename}
	 }
	 \fi

\end{document}